\documentclass[%
showpacs,preprintnumbers,
 amsmath,amssymb,
 aps,
 pre,
 lengthcheck,%
]{revtex4-1}
\usepackage{graphicx}
\usepackage{dcolumn}
\usepackage{bm}
\usepackage[latin1]{inputenc}
\usepackage{color}
\usepackage{amssymb}
\usepackage{amsmath}
\usepackage{mathrsfs}
\usepackage{multirow}
\usepackage{booktabs}

\begin{document}


\title{Lattice splitting under intermittent flows}


\author{Markus Schl\"apfer and Konstantinos Trantopoulos}
\affiliation{Laboratory for Safety Analysis, ETH Zurich, 8092 Zurich, Switzerland}


\date{\today}

\begin{abstract}
We study the splitting of regular square lattices subject to stochastic intermittent flows. Various flow patterns are produced by different groupings of the nodes, based on their random alternation between two possible states. The resulting flows on the lattices decrease with the number of groups according to a power law. By Monte Carlo simulations we reveal how the time span until the occurrence of a splitting depends on the flow patterns. Increasing the flow fluctuation frequency shortens this time span which reaches a minimum before rising again due to inertia effects incorporated in the model. The size of the largest connected component after the splitting is rather independent of the flow fluctuation frequency but slightly decreases with the link capacities. Our findings carry important implications for real-world networks, such as electric power grids with a large share of renewable intermittent energy sources.
\end{abstract}


\maketitle

\section{Introduction} \label{intro}
Assessing the robustness of networks against failures of nodes and links is an essential research topic across many scientific disciplines. Examples range from the extinction of species in food webs and malfunctions in protein networks to the vulnerability of the World Wide Web and cascading failures in electric power grids. In the last decade, substantial new insights have been gained through the application of methods from statistical physics \cite{Newman:2003,Boccaletti:2006, Barratt:2008, Doro:2008}. Random failures as well as targeted attacks have been addressed by first studying static properties such as different network topologies \cite{albert2000}. Later on, load redistribution models have been introduced to better represent networks supporting the flow of a physical quantity. For example, the load of a node has been defined by its betweenness centrality \cite{Motter:2004}, by the total number of efficient paths passing through it \cite{Crucitti:2004}, or enriched with stochastic flux fluctuations \cite{Heide:2008}. While these approaches model the failure propagation in a static manner, the dynamic flow properties have just recently been taken into account \cite{Lubos:2008}. 
  
The contribution of this paper is to investigate the impact of stochastic intermittent flow patterns on the potential occurrence of cascading link failures, eventually leading to a network splitting. Therefore, our model considers 2-dimensional lattices with different groups of nodes which randomly alternate between two possible states, i.e. they act as sources or sinks respectively. These state transitions induce time-varying stochastic flows on every link. Once reaching its capacity, a link fails with a time delay due to inertia effects. 

The motivation for this dynamic flow model was the large-scale integration of renewable intermittent energy sources (e.g. wind power, photovoltaic systems) into the electric power grid. This implies a higher ratio of non-dispatchable generation which, in turn, leads to less predictable and more fluctuating flows on the network. Consequently, the anticipation of undesired situations such as cascading transmission line overloads leading to a network breakdown becomes highly complicated \cite{papa:2006}. In such a future infrastructure layout the network merely serves as a backbone for the redistribution of power from regions of energy surplus to regions with net power consumption. As detailed modeling and simulation approaches become limited due to the increased complexity of electric power systems with large share of renewables, we opted for a minimalistic approach in order to understand the fundamental physics governing the dynamic behavior leading to a network splitting. Past experience has shown that such a splitting potentially results in a wide-area blackout with severe social and economic consequences \cite{Usbo:2004}.

Questions to be tackled are: What is the relation between the stochastic behavior of the nodes and the emerging flow patterns on the network? How are these flow patterns affected by different groupings of the nodes? What, in turn, is the impact of these flow patterns on the probability of a network splitting? How do inertia effects influence the potential splitting process? 

Although the definition of our model is based on the specific properties of future energy networks, it is expected to reflect basic features of other real-world networked systems, whose robustness is subject to stochastic intermittent flows. 

\section{Dynamic Model and Simulation Procedure}
Our study system incorporates a model for the nodal state alternation, a flow model, a lattice layout model and a model for the cascading link outages.
\subsection{Stochastic nodal state alternation}
The two possible states between which all nodes can alternate assume current injections of $P_i^{+}=1$ (node state ``up'') when the node acts as a source or $P_i^{-}=-1$ (node state ``down'') when the node acts as a sink. This stochastic up-down-up cycle assumes for every node $i$ constant transition rates $\lambda_i$ and $\mu_i$ respectively. Hence, this alternating process is characterized by the cumulative distribution functions of the up-state and down-state times,
\begin{equation}\label{eq:GenF}
F_i(t_u) = 1-e^{-\lambda_i t_u} , \qquad F_i(t_d) = 1-e^{-\mu_i t_d},  
\end{equation}
where $t_u$ and $t_d$ are the time spans measured from the moment of entering the up-state and down-state respectively. The state transition frequency of every node is calculated by
\begin{equation}\label{eq:renewalDensity}
f_i =  \frac{\lambda_i \mu_i}{\lambda_i + \mu_i}
\end{equation}
and corresponds to the average number of up-down-up cycles per time unit. For simplicity we assign to every node $i$ the same transition rates $\lambda_i=\lambda$ and $\mu_i=\mu$, implying the same transition frequency $f_i=f$. Moreover, the ratio is kept constant at $\lambda/\mu=1$ in order to assign the same probabilities to both possible states.
\subsection{Flow model}
We model the flows on the network by applying an electrical direct current model based on Ohm's law. Thereby, the linear relation between the nodal current injections $P_i$ and the voltages $V_i$ can be put into matrix form
\begin{equation}\label{Eq:flow}
\mathbf{P} = \mathbf{B} \mathbf{V}.
\end{equation}
The conductance matrix $\mathbf{B}$ has elements $B_{ij}=-r_{ij}^{-1}$ and $B_{ii}=\sum_{j \in \Omega_i}r_{ij}^{-1}$ where $r_{ij}$ is the resistance of each link $(i,j)$ and $\Omega_i$ is the set of all the directly connected nodes to $i$. By assuming for simplicity that $r_{ij}=1$ for all links, the flow on a link $(i,j)$ is given by
\begin{equation}\label{Eq:flow_ij}
P_{ij}=V_i-V_j.
\end{equation}
The sum of all the current injections at a given time instant is not necessarily equal to zero due to the stochastic nature of the up-down-up cycle. In order to satisfy the balance condition $\sum_i P_i = 0$ at all times, a lack or surplus of the total current injections within the network is compensated by an additional, equally distributed injection $\pm |({\sum_i P_i})/N |$ at every node. Nevertheless, the satisfaction of the balance condition implies that the rows of $\mathbf{B}$ are linearly dependent. To make Eq. (\ref{Eq:flow}) uniquely solvable, one of the equations in the system is removed and the node associated with that row is chosen as the voltage reference $V_{ref}=0$. 
\subsection{Lattice layout and node grouping}
\begin{figure*}
\centering
\includegraphics{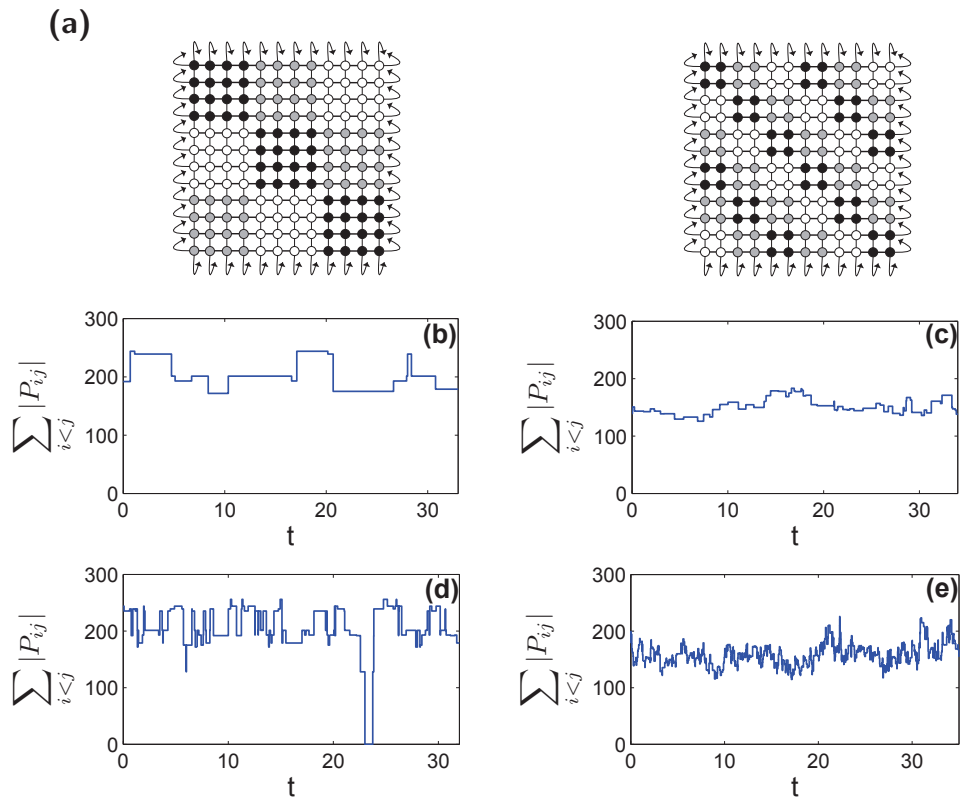}
\caption{\label{Fig:1}(Color online) (a) Schematic plot of a $12\times12$ lattice ($N=144$) with grouping factor $G=9$ (left) and $G=36$ (right). Note that the coloring of the groups has been used only due to illustrative reasons, all groups are stochastically independent with each other. The total flow $\sum_{i < j}|P_{ij}|$ versus time in the left lattice is depicted in (b) for $f=0.01$ and in (d) for $f=0.5$. Similar for the right lattice in (c) and (e) respectively. Notice that an increased $f$ results in a higher fluctuation frequency of the flows. By further varying the grouping factor $G$, a broad spectrum of different stochastic flow patterns can be reproduced. A high value of $G$ is leading to more smooth flow time-series, while a small value implies a strong fluctuation around the mean value.}    
\end{figure*}

We embedded our model for the nodal behavior and the resulting flows in a regular square lattice of $N$ nodes and $L=2N$ links with periodic (or ``wrap-around'') boundary conditions. In this way every node is directly connected to 4 neighbors, thus different conditions for boundary nodes are avoided. Furthermore, we partition the lattice into several square groups, each containing an equal number of nodes [Fig. \ref{Fig:1} (a)]. All the nodes in a given group are in the same state at all times and alternate states simultaneously. We denote the grouping factor $G$ as the number of groups in the network, thus $G=N$ represents total stochastic independence between all nodes. As depicted in Fig. \ref{Fig:1} (b)-(e), an increased $f$ results in a higher fluctuation frequency of the flows. By further varying the grouping factor $G$, a broad spectrum of different stochastic flow patterns can be reproduced. A high value of $G$ is leading to more smooth flow time-series, while a small value implies a strong fluctuation around the mean value.
\subsection{Link outage model}
In order to incorporate inertia effects in our model, the link outage mechanism is based on the concept that the flow $P_{ij} (t)$ determines the ``temperature'' $T_{ij}(t)$ on the link $(i,j)$ according to  
\begin{equation}\label{eq:tempDiffEq}
\tau_{ij}\frac{dT_{ij} (t)}{dt} = q_{ij} P_{ij} (t) -  T_{ij} (t).
\end{equation}
The link fails if $T_{ij}(t)$ reaches its capacity $T_{ij}^{c}$. In order to simplify Eq. (\ref{eq:tempDiffEq}) we set $q_{ij} = 1$. The parameter $\tau_{ij}$ represents the characteristic time (inertia) constant. 

As an example, such an inertia is present in electric power grids where the power flows might heat the transmission lines up to a maximum allowable temperature. 
\subsection{Simulation procedure}
With respect to the implementation we opted for a discrete-event based approach. This allows describing the time evolution of the nodal states and the resulting flows, as well as of the link outages and the resulting lattice status. By means of extensive Monte Carlo simulations we estimated the expected time until the splitting of the lattice.
The simulation procedure comprises the following steps:
\begin{enumerate}
\item 
Construct the $N\times N$ lattice adjacency matrix $\mathbf{A}$ and the $N\times N$ conductance matrix $\mathbf{B}$. For all the nodes $i$ in a single group determine their equal output states $P_i$ at $t=0$ by a single Bernoulli trial with probability $p=0.5$. Set the simulation step to $n=0$, set $t_{(0)}=0$ and initialize the temperature of each link to $T_{ij}(t_{(0)})=0$.\\ 
\item
Calculate the flow $P_{ij}$ on each link $(i,j)$ by \mbox{Eq. (\ref{Eq:flow_ij})} after solving Eq. (\ref{Eq:flow}) for $\mathbf{V}$.
For all links $(i,j)$ determine the subsequent time step \mbox{$\Delta t^{temp}_{ij,(n+1)}$} after which they fail. If $P_{ij}(t_{n})\geq  T_{ij}^{c}$, this time span is given by
\begin{equation}\label{eq:deltaTBeta}
\Delta t_{ij,(n+1)}^{temp} = -\tau_{ij} \ln \bigg( \frac{T_{ij}^{c} - P_{ij}(t_{(n)})}{T_{ij}(t_{(n)})-P_{ij}(t_{(n)})}\bigg).
\end{equation} 
\\
For every link calculate the point in time when it fails due to reaching $T_{ij}^{c}$ as $t^{temp}_{ij} = t_{(n)}+ \Delta t_{ij,(n+1)}^{temp}$ and build the vector $t^{temp}$ with elements $t^{temp}_{ij}$. Determine the time of the first link outage as $t^{out,temp}=\min[t^{temp}]$.
Determine for every node $i$ of the network the point in time $t_{i}^{s}$ when it changes its state. Then, the time of the first state change is given by \mbox{$t^{change,s}= \min[t^s]$ with

$t^s =\begin{array}{cccc}[t_{1}^s&  t_{2}^s& \cdots & t_{N}^s]\end{array}$}.
Determine the time of the next simulation event as $t^{next}= \min[t^{out,temp},t^{change,s}]$. Increment the simulation step to $n=1$. 
\item 
Proceed the simulation to $t_{(n)}=t^{next}$. Remove the failed link (if any) from the lattice and update $\mathbf{A}$ and $\mathbf{B}$. Recalculate the output $P_i$ of each node $i$ based on Eq. (\ref{eq:GenF}).
Recalculate the flow $P_{ij}$ and the temperature $T_{ij}$ on each link $(i,j)$. The flow $P_{ij}$ remains constant at least until the next event. The temperature $T_{ij}(t_{(n)})$ is given by
\begin{eqnarray}\label{eq:lossB}
T_{ij}(t_{(n)}) & = & P_{ij}(t_{(n-1)}) \Big[1-e^{-\frac{1}{\tau_{ij}}\Delta t_{(n)}} \Big] {} \nonumber\\
\vspace{2cm} &&  +T_{ij}(t_{(n-1)}) e^{-\frac{1}{\tau_{ij}}\Delta t_{(n)}},
\end{eqnarray}
where $\Delta t_{(n)}=t_{(n)}-t_{(n-1)}$.
\item For each node and link recalculate $t_{i}^{s}$ and $t^{temp}_{ij}$ and update $t^{change,s}$ and $t^{out,temp}$ respectively as described in Step 2. Determine the time of the next event $t^{next}$. 
\item Check the connectivity of the lattice. If it remains connected, increment the simulation step $n$ and go back to Step 3. Otherwise, stop the simulation.
\end{enumerate}
\section{Numerical results}
\subsection{Average flows}
\begin{figure}
\includegraphics[scale=0.5]{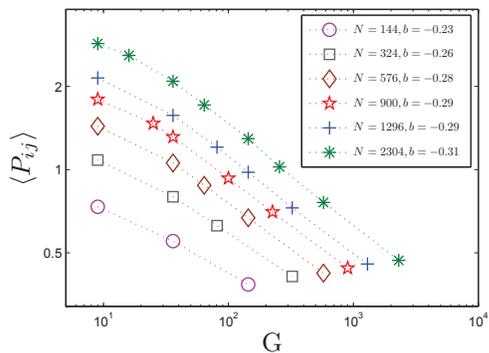}
\caption{\label{fig:avflows}(Color online) Average flows $\left\langle P_{ij} \right\rangle$ versus the grouping factor $G$, which are well fitted by a power law with characteristic exponents $b$, slightly increasing with the lattice size $N$. The dotted lines serve as a guide to the eye.}
\end{figure}

To clarify the impact of different grouping factors $G$ and lattice sizes $N$ on the resulting flow patterns, we estimate the average flow per link $\left\langle P_{ij} \right\rangle$ without considering the link outage model,
{\setlength\arraycolsep{4pt}
\begin{eqnarray}\label{eq:avFlow}
\left\langle P_{ij} \right\rangle & = & \frac{1}{L}\lim_{t \to \infty} \bigg(\frac{1}{t} \int_0^{t} \sum_{i<j}|P_{ij}(t')|dt'\bigg){} \nonumber\\
\vspace{2cm} &  \approx & \frac{1}{t^{tot}L} \sum_{n} \sum_{i<j}\big(|P_{ij}(t_{(n-1)})|\Delta t_{(n)}\big),
\end{eqnarray}}
where $t^{tot}= \sum_n {\Delta t_n}$ is the sampled overall time span. The average flow is independent of $f$, as by increasing the frequency the relative duration among all different nodal state combinations remains unchanged. Figure \ref{fig:avflows} shows the values of $\left\langle P_{ij} \right\rangle$ versus the grouping factor $G$ in lattices of different sizes $N$.
\begin{figure}
\includegraphics[scale=0.45]{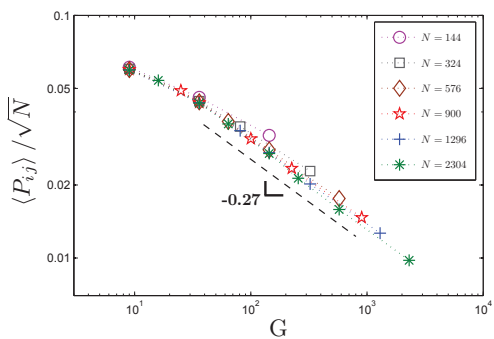}
\caption{\label{fig:avflowsScale}(Color online) Collapse of all the average flow data shown in Fig. \ref{fig:avflows} by scaling as $\left\langle P_{ij} \right\rangle / \sqrt{N}$, versus the grouping factor $G$. The collapsed data follow a power law with characteristic exponent $-0.27$. The dotted lines serve as a guide to the eye.}
\end{figure}
The average flows increase with the lattice size because for a given $G$ the number of nodes in a group increases with $N$ leading to a higher current exchange among the groups. Moreover, $\left\langle P_{ij} \right\rangle$ decreases with the number of independently alternating groups. For a given lattice size $N$, increasing $G$ implies less nodes in the groups and thus less exchange among them. Seen from a different angle, a high value of $G$ means that less nodes behave simultaneously in the same way, leading to a more local current exchange and less flows in the lattice. In contrast, a low $G$ induces higher flows over longer distances. Interestingly, for a given $N$ the decrease of the average flow with $G$ follows a power law
\begin{equation}\label{eq:avFlowsDecay}
\left\langle P_{ij} \right\rangle \propto G^{-b}, G\geq9.
\end{equation}
The exponent $b$ is rather small and slightly increasing with the size of the lattice. 

As shown in Fig. \ref{fig:avflowsScale} the data in Fig. \ref{fig:avflows} collapse onto a single curve, if the average flows are scaled with $\sqrt{N}$.
This result can be explained by the flow distribution on the lattice. The average flow is largely determined by the maximum flows which are encountered at the boundaries of the groups. Suppose two lattices with sizes $N_1$ and $N_2$ and same grouping factor $G$. Then the maximum possible flows induced by a (square) group on one of its boundary links, $P_1^{max}$ and $P_2^{max}$, are approximately proportional to $\sqrt{N_1/G}$ and $\sqrt{N_2/G}$ with the same factor respectively. Thus $P_1^{max}/P_2^{max} \approx \sqrt{N_1/N_2}$.

\subsection{Lattice splitting}
The robustness of a lattice is quantified by the expected time $\left\langle t_{split} \right\rangle$ when a splitting occurs and the lattice breaks into two parts \footnote{Notice that the cascading link outages might propagate further, eventually leading to a lattice split in more than two components. Nevertheless, we stop the simulation at this point as our study is mainly motivated by electric power grids where a first splitting might already lead to a complete system breakdown (blackout) due to instability phenomena.}. 
\begin{figure}[b!]
\centering
\resizebox{0.75\columnwidth}{!}{%
\includegraphics{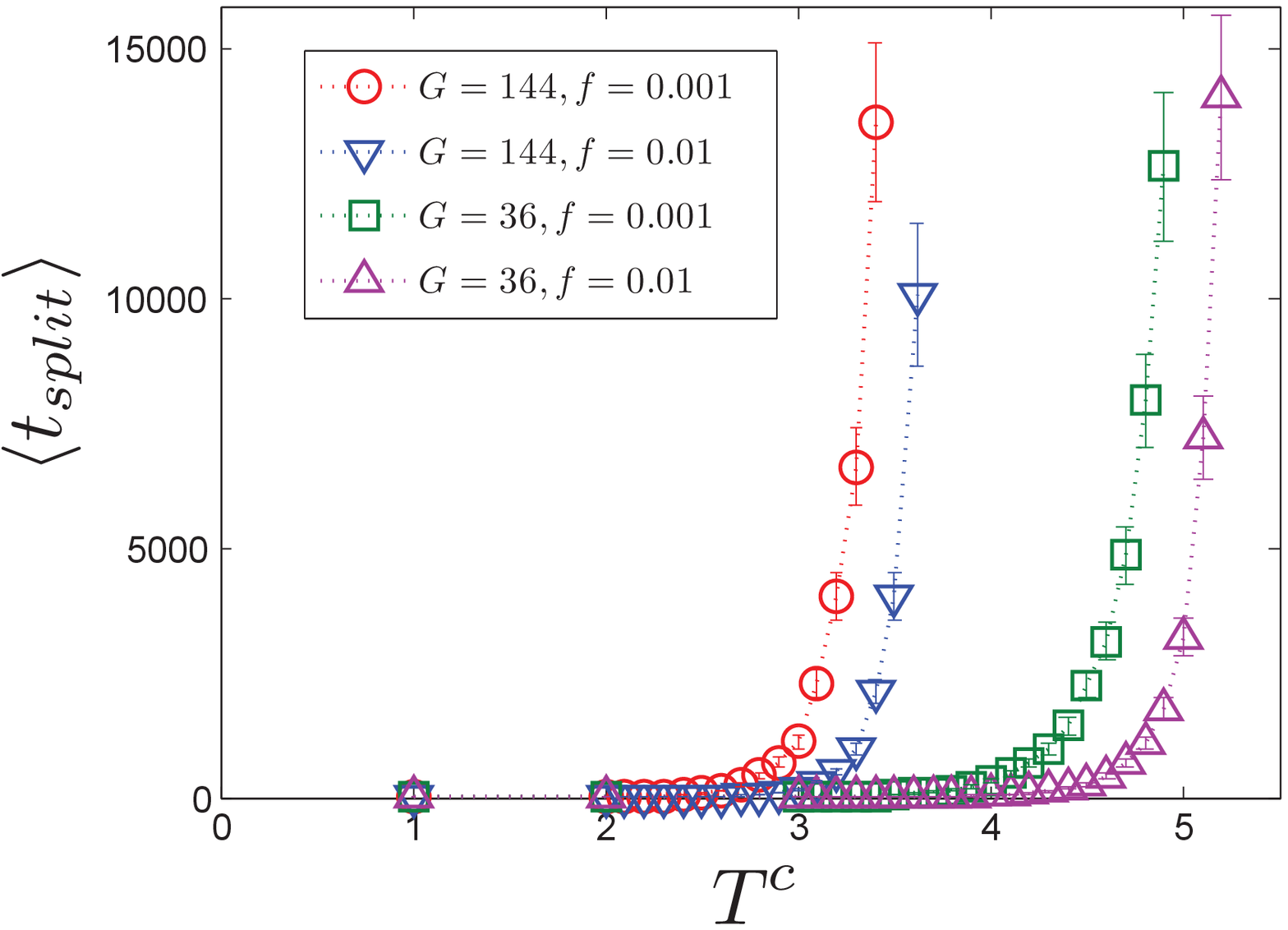}}
\caption{\label{fig:Tc}(Color online) Relationship between the link capacity $T^c$ and the expected time until the splitting, $\left\langle t_{split} \right\rangle$, of a $24 \times 24$ lattice ($N=576$) for two different grouping factors $G$ and state transition frequencies $f$. The inertia constant is set to $\tau=1$. The error bars indicate the 95\% confidence interval. The dotted lines serve as a guide to the eye.}
\end{figure}
\begin{figure*}
\centering
\resizebox{1.4\columnwidth}{!}{%
\includegraphics{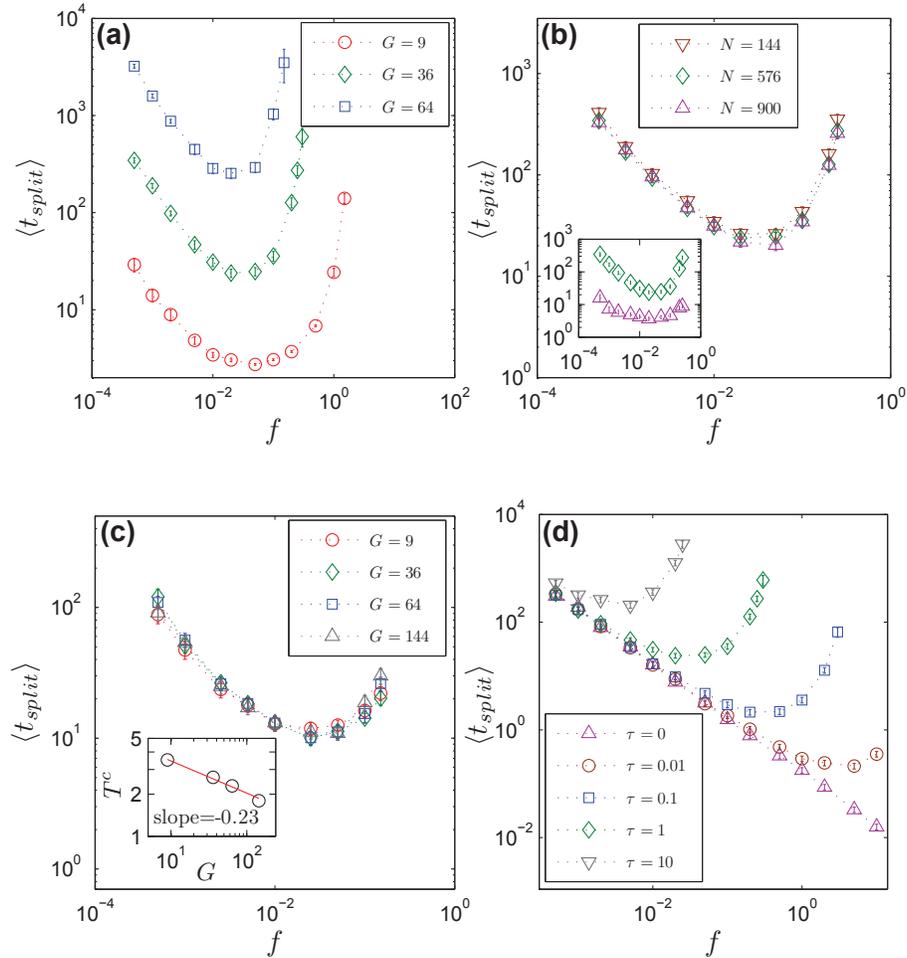}}
\caption{\label{fig:frequency}(Color online) Expected time $\left\langle t_{split} \right\rangle$ until the lattice splitting versus the state transition frequency $f$. If not stated otherwise, the lattices have size $N=576$. The error bars indicate the 95\% confidence interval. (a) Effect of different grouping factors $G$. The parameters of the dynamic model are set to $\tau=1$ and $T^c=4$. (b) Collapse of the splitting time data in lattices of different sizes $N$ with $G=36$, by adjusting the link capacities according to $T^c = a \sqrt{N}$ with $a=1/6$. The inertia constant is set to $\tau=1$. The average splitting times without adjusting $T^c$ are depicted in the inset for $N=576$ and $N=900$. The corresponding values for $N=144$ are omitted as the high splitting times induce prohibiting simulation run-times. (c) Collapse of $\left\langle t_{split} \right\rangle$  with $\tau=1$ due to adjusting $T^c$ according to the grouping factor $G$. The inset shows the chosen value of $T^c$ for each value of $G$, being fitted by a power law. (d) Effect of the inertia constant $\tau$ on the splitting times with $G=36$ and $T^c=4$.}
\end{figure*}
This time span can be interpreted as the life expectancy of the lattice and depends on the capacity $T_{ij}^c$ of each link $(i,j)$. To simplify matters, we assign the same value $T_{ij}^c=T^c$ and $\tau_{ij}=\tau$ to all links. Figure \ref{fig:Tc} shows the behavior of $\left\langle t_{split} \right\rangle$ versus an increasing value of $T^c$ for two different grouping factors $G$ and state transition frequencies $f$.
The expected time until the lattice splits increases exponentially with the link capacity. Hence, $\left\langle t_{split} \right\rangle$ is highly sensitive with respect to small changes of $T^c$. For a given value of $T^c$, a larger grouping factor $G$ leads to a significantly higher value of $\left\langle t_{split} \right\rangle$, as less flows are induced [Fig. \ref{fig:avflows}]. The effect of varying the state transition frequency $f$ is similarly large and is examined in more detail in Fig. \ref{fig:frequency}. Starting with a low value, an increase of $f$ leads to a shorter time span until the combined nodal states induce those minimum flows which are needed for the temperatures $T_{ij}$ to reach the capacities $T^c$ [Fig. \ref{Fig:1}]. Consequently, as depicted in Fig. \ref{fig:frequency} (a)-(d), the splitting times $\left\langle t_{split} \right\rangle$ are high at low values of $f$ and become significantly decreased as the value of $f$ is increasing. However, as $f$ is exceeding a certain value, the splitting times start to increase again, and the lattice becomes more robust. This result can be explained by the inertia effects according to Eq. (\ref{eq:tempDiffEq}). While the flows on the lattice reach more often higher absolute values [Fig. \ref{Fig:1}], the average residence times of the underlying nodal states begin to fall below the minimum time needed to heat the links up to their capacity $T^c$. 

With a small number of groups the combined output is more fluctuating between the extreme values [Fig. \ref{Fig:1}, left, compared to Fig. \ref{Fig:1}, right)]. This, in turn, is leading to a higher probability to encounter high flows and high link temperatures in a given time span. The splitting times $\left\langle t_{split} \right\rangle$ thus are significantly shorter, as depicted in Fig. \ref{fig:frequency} (a). 

By considering the scaling behavior of the average flows [Fig. \ref{fig:avflowsScale}], the values of $\left\langle t_{split} \right\rangle$ collapse for lattices with different sizes, but equal model parameters otherwise, if the link capacities are set as $T^c \propto \sqrt{N}$. This result is demonstrated in Fig. \ref{fig:frequency} (b) for three different lattice sizes. Notice that for a given value of $f$ the average splitting times without adjusting $T^c$ differ by several decades [Fig. \ref{fig:frequency}, inset].

As shown in Fig. \ref{fig:frequency} (c) the link capacities $T^c$ can be adjusted in such a way, that the splitting times in lattices with equal $N$ but different grouping factors $G$ overlap for a wide range of the state transition frequency $f$. The adjusted capacities can be fitted by a power law with characteristic exponent $-0.23$ [Fig. \ref{fig:frequency} (c), inset], being remarkably close to the characteristic exponent $b$ of \mbox{Eq. (\ref{eq:avFlowsDecay})}.

The effect of the inertia is shown in Fig. \ref{fig:frequency} (d) by varying the inertia constant $\tau$. Without any inertia, i.e. $\tau=0$ implying $T_{ij}(t)=P_{ij}(t)$ [Eq. (\ref{eq:tempDiffEq})], the splitting time declines with slope $1/f$. In average, the number of state transition events increases linearly with $f$ in a given time span. This, in turn, decreases the average time until a maximum allowable flow $P_{ij}=T^c$ on a link $(i,j)$ is reached, in an inversely proportional manner. However, for $\tau>0$ a minimum average splitting time arises for a roughly estimated value of $f \approx 0.05 /\tau$.

\begin{figure}
\centering
\resizebox{0.9\columnwidth}{!}{%
\includegraphics{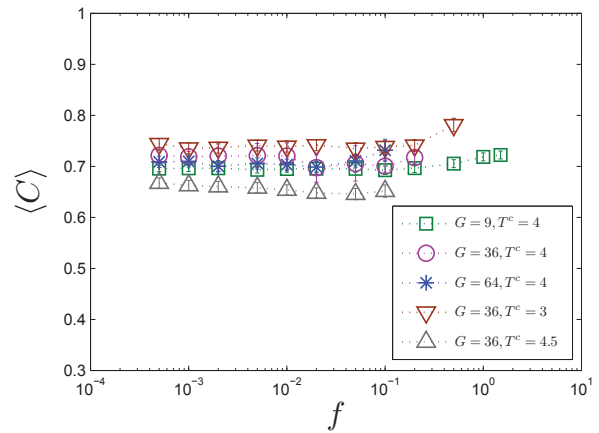}}
\caption{\label{fig:component}(Color online) Average size of the larger connected component $\left\langle C \right\rangle$ after the splitting into two parts versus the state transition frequency $f$ for various grouping factors $G$ and link capacities $T^c$. The lattices have size $N=576$. The error bars indicate the 95\% confidence interval. The dotted lines serve as a guide for the eye.}
\end{figure}

In order to quantify the damage after the splitting, we follow \cite{Motter:2002} and measure the average relative size of the larger of the two remaining connected components
\begin{equation}\label{eq:avFlows}
\left\langle C \right\rangle=\left\langle N'\right\rangle/N,
\end{equation}
where $\left\langle N'\right\rangle$ denotes the average number of nodes in the larger connected component. Figure \ref{fig:component} shows the average relative size of the larger connected component $\left\langle C \right\rangle$ for different grouping factors $G$ and link capacities $T^c$ versus the state change frequency $f$.
The size of the larger connected component is rather independent of the flow fluctuation frequency as determined by $f$. While keeping the same link capacities [Fig. \ref{fig:component}, values for $T^c=4$] there is no clear indication with regard to the dependence of $\left\langle C \right\rangle$ on the grouping factor $G$. For a given $G$, decreasing the link capacities $T^c$ slightly increases the value of $\left\langle C \right\rangle$. For a smaller value of $T^c$ lower flows are sufficient to overload the links and cascades may develop in smaller regions. Hence, the failing links envelop a lower number of nodes eventually breaking away from the lattice, implying a larger size of the remaining connected component after the splitting. However, for the chosen values of $T^c$ the average relative size remains approximately in the range $0.65 < \left\langle C \right\rangle < 0.75$. 
\section{Summary and Conclusions}
To sum up, in this paper we have introduced a minimal model for stochastic intermittent flows on lattices. These flows might induce cascading link overloads eventually leading to a lattice splitting into two parts. In order to better represent real systems we implied an inertia in such a way that a link does not fail immediately but rather delayed when it becomes overloaded. By extensive Monte Carlo simulations we revealed how the time until such a splitting occurs depends on different flow patterns. With an increasing number of (stochastically) independent nodes the average flows decrease slowly, following a power law. With regard to the robustness of the lattices, a high sensitivity of the splitting time to the link capacities is observed. Increasing the flow fluctuation frequency (as determined by the nodal state alternation) decreases this time span until reaching a minimum, after which it rises again meaning a higher ``life expectancy'' of the lattice. Generally, both a higher stochastic independence among the nodes (i.e. more groups of simultaneously alternating nodes) and a smaller size of the lattice imply higher splitting times. However, these time spans seem to coincide by adjusting the link capacities according to a power law with respect to the node grouping, and according to the square-root of the lattice size respectively. Furthermore, we have shown that the effect of the inertia is significant. Its absence implies a monotonic decrease of the splitting times, while introducing it results in remarkably higher values for higher inertia constants. As an indication of the damage after the splitting, the relative size of the larger connected component seems to be independent of the flow fluctuation frequency but sligthly decreases with the link capacity.

We conclude with some thoughts on the implications of these results for future energy networks, being characterized by a large share of renewable intermittent power sources. The more distributed the power sources are (being equivalent to more groups in our model), the lower the flows exchanged over the power grid [Fig. \ref{fig:avflows} and Fig. \ref{fig:frequency} (a)]. However, even in a highly distributed system, a considerable transmission capacity is still needed to keep the system at the desired level of security [Fig. \ref{fig:frequency} (c)]. Increasing the size of the grid can be expected leading to a disproportionally small increase of the flows [Fig. \ref{fig:avflowsScale}]. Restricting the capacities of the transmission lines or, equivalently, operating the system closer to its security margins might reduce the robustness of the network against cascading failures drastically [Fig. \ref{fig:Tc}]. The inertia as induced by the heating of the transmission lines which might fail when reaching a maximum allowable temperature, potentially increases the robustness of power grids with large share of renewables [Fig. \ref{fig:frequency} (d)]. The same effect can be even exploited for increasing existing transmission line capacities, thus improving the economic performance of the system \cite{Schlaepfer:2010}. If the grid breaks apart as a result of cascading line failures, the sizes of the two formed islands can be expected to be largely independent of the flow fluctuation frequencies. Nevertheless, they seem to slightly become more asymmetric with decreasing power transfer capacities thus leaving a larger remaining connected component of the network [Fig. \ref{fig:component}].

Our model provides insights into the underlying physics of networks subject to stochastic flows. Therefore, we believe that besides future energy networks, potential applications could be investigated on other real-world systems such as traffic networks.
\begin{acknowledgments}
We wish to thank Wolfgang Kr\"oger, Paul Expert, Giovanni Petri, Henrik Jeldtoft Jensen, Kyriakie Kyriaki and Sven Dietz for helpful discussions and encouragement. M.S. acknowledges ``swiss\emph{electric} research'' for co-funding the present work. K.T. acknowledges partial financial support by the Swiss Federal Office for Civil Protection.   
\end{acknowledgments}


%

\end{document}